# Sculpting the plasmonic responses of nanoparticles by directed electron beam irradiation

Kevin M. Roccapriore[1,*], Shin-Hum Cho[2,†], Andrew R. Lupini[1], Delia J. Milliron[2], and Sergei V. Kalinin[1,*]

[1] Center for Nanophase Materials Sciences, Oak Ridge National Laboratory, Oak Ridge, TN 37831

[2] McKetta Department of Chemical Engineering, The University of Texas at Austin, Austin TX 78712 USA

**Abstract**

Spatial confinement of matter in functional nanostructures has propelled these systems to the forefront of nanoscience, both as a playground for exotic physics and quantum phenomena and in multiple applications including plasmonics, optoelectronics, and sensing. In parallel, the emergence of monochromated electron energy loss spectroscopy (EELS) has enabled exploration of local nanoplasmonic functionalities within single nanoparticles and the collective response of nanoparticle assemblies, providing deep insight into the associated mechanisms. However, modern synthesis processes for plasmonic nanostructures are often limited in the types of accessible geometry and materials, and even then, limited to spatial precisions on the order of tens of nm, precluding the direct exploration of critical aspects of the structure-property relationships. Here, we use the atomic-sized probe of the scanning transmission electron microscope (STEM) to perform precise sculpting and design of nanoparticle configurations. Furthermore, using low-loss (EELS), we provide dynamic analyses of evolution of the plasmonic response during the sculpting process. We show that within self-assembled systems of nanoparticles, individual nanoparticles can be selectively removed, reshaped, or arbitrarily patterned with nanometer-level resolution, effectively modifying the plasmonic response in both space and energy domains. This process significantly increases the scope for design possibilities and presents opportunities for arbitrary structure development, which are ultimately key for nanophotonic design. Nanosculpting introduces yet another capability to the electron microscope.

[†] Present address: Samsung Electronics, Samsung Semiconductor R&D, Hwaseong, Gyeonggi-do 18448, Republic of Korea

**Introduction**

Spatial confinement of matter in functional nanoparticles has propelled these systems to the forefront of a variety of fields in nanoscience and nanotechnology.[1–4] Nanoparticles offer an excellent environment for exploration of fundamental physical and chemical phenomena due to their extreme degrees of tunability.[5–7] As a result, they have been applied in a variety of fields ranging from nanophotonics, quantum optics, catalysis, medicine, mechanical or optical coatings, to surface enhanced Raman spectroscopy (SERS).[8–13]

Nanoparticles on their own or in few-particle clusters exhibit properties not found in the bulk due to a large ratio of surface area to volume; however, controlled ordering of nanoparticles assemblies allows development of new functionalities, leading to concepts such as single molecule sensing,[14] artificial molecules,[15] and molecular imaging.[16] This in turn leads to the dual challenge of preparation of nanoparticle assemblies with desired geometries and exploration of their functionalities of interest. An enormous variety of methods exist for nanoparticle synthesis including chemical synthesis and physical vapor deposition.[17–22] Electron beam lithography (EBL), focused ion beam (FIB) milling, and laser-induced dewetting methods can be used for control over material positioning and growth to some degree, but resulting particle sizes are currently too large for applications requiring particles with less than about 20 nm as the critical dimension. Colloidal self-assembly of nanocrystals is a chemical growth alternative that allows for high packing density and high quality nanoparticle ordering for particle sizes below 10 nm and has been thoroughly reviewed,[23] but in general, the particle chemistry, stoichiometry, morphology and relative position largely cannot be locally modified after the assembly process, which hinders opportunities to explore phenomena that depend on these traits.

Perhaps the best way to extend the correlations between structure and properties is through a combination of these methods via controllable modification of synthesized plasmonic material. Intentionally modifying the plasmonic structures dynamically after synthesis has largely not been explored. Direct-write techniques such as EBL and FIB have high degrees of precision, and can be used for patterning down to the ~1 nm level with an aberration corrected beam in specific geometries.[24] An even higher-precision alternative is the electron beam of the scanning transmission electron microscope (STEM). It is well known that the beam can modify the material, often in an undesirable[25] manner, however the beam-sample interaction can also be exploited to directly control and manipulate materials down to the single-atom level.[26–29] The electron beam can also be used to create larger scale nanostructures, for example by milling nanowires to tune plasmonic resonances.[30]

From a characterization perspective, the breakthrough in investigating plasmonic behavior of individual and few nanoparticle systems resulted from advances in monochromated electron energy loss spectroscopy (EELS) in an aberration corrected STEM. This route has enabled direct visualization of the spatial distribution of plasmon resonances in nanocrystals,[31] helped discover that detected resonances in quantum nanoparticles were different than original theoretical predictions,[32] and more recently, allowed distinguishing between different elemental isotopes via so-called aloof spectroscopy.[33] These experiments, however, rely on pre-existing as-synthesized nanoparticle geometries and thus only a limited degree of exploration is possible.

Here we demonstrate the *in-situ* control of plasmonic structures within the framework of a self-assembled monolayer of nanoparticles using an electron beam in a STEM, where the electron beam is accelerated to 60 kV and is aberration corrected up to $5^{th}$ order. This method allows us to explore near-arbitrary structures, dynamically tune plasmon resonances, and examine the chemical and plasmonic evolution during *in-situ* nanoparticle manipulation using EELS.

The material being studied is a self-assembled monolayer of chemically synthesized $F,Sn:In_2O_3$ semiconductor nanoparticles with a typical dimension of 10 nm. These nanoparticles offer an ideal platform for spectrally tuning plasmonic responses based on electron donation from Sn and F dopants[34] that are spectrally far from the interband transition. Therefore, plasmon modes can be unambiguously identified. The STEM used to sculpt the nanoparticles is also used to analyze the modifications by EELS, allowing a fast and convenient testing ground for both chemical analysis via acquiring core loss energy signals in real time, and plasmonic analysis via low-loss energies. Specifically, the modification and *in-situ* core-loss EELS are carried out at high current, while the (monochromated) low-loss EELS analysis is conducted at roughly 20x lower beam current.

A typical high angle annular dark field (HAADF)-STEM image of a self-assembled monolayer of nanocubes is shown in **Figure 1** (a), highlighting that, even in a close to best-case scenario, a large degree of heterogeneity and randomness is present in the system. Since the particle sizes fluctuate around a nominal average of 10 nm, size-dependent effects may play a significant role in the plasmonic response, such that slight deviations in particle size have a strong impact.[35,36] The nanoparticle array itself supports several plasmon modes, most of which are hybridized modes and difficult to analyze due to their collective nature and spatial overlap[37]. However, when a defect in the array (e.g., an edge or a missing nanoparticle) is encountered, the electric field strength and thus the plasmonic response is greater. This is important for many applications that require an enhanced field strength in a specific location that, as we will show, can be achieved by manipulation of a nanoparticle's position using the electron beam.

Several hallmark examples of nanoparticle modifications are presented in Figure 1. First, we generate features by removing other particles surrounding the intended feature; Figure 1(b) shows a 1-D chain; Figure 1(c) shows a split-ring resonator – an element useful in nanophotonics and metamaterials. Different checkerboard patterns illustrate periodic plasmonic hotspots, Figures 1(e) and (f). Alternatively, particles may be fused together under different scanning schemes (Figure 1(g)). The use of the electron beam in these ways allows tremendous flexibility in post-fabrication and enables the ability to tune the location of nanoparticles as well as their plasmonic spatial features. While we measure changes in plasmonic properties, we also expect other useful changes to accompany the modifications, which could hold the potential for applications in catalysis or energy-harvesting fields.

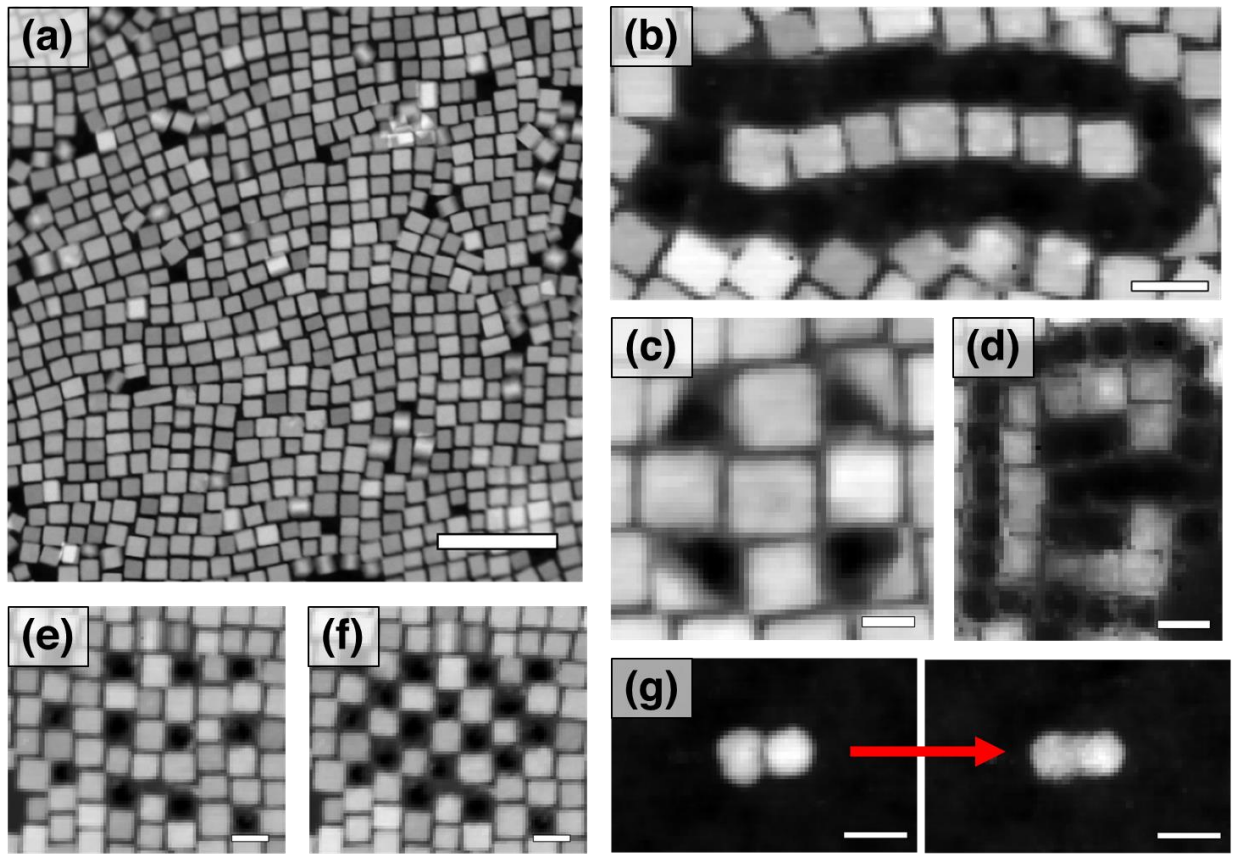

**Figure 1**. Examples of structural modification to nanoparticles by electron beam irradiation, visualized by HAADF-STEM images. (a) As-assembled monolayer; (b) a 1D nanochain carved out from a large array; (c) nanoparticles can be sliced diagonally or arbitrarily; (d) split ring resonator element, critical in plasmonics and metamaterial designs, constructed of nanoparticles; (e) and (f) depict various checkerboard-type schemes; (g) fusing of two isolated nanoparticles into a single nanoparticle. Scale bar in (a) is 100 nm, (c) 10 nm, all others 20 nm.

As a rough estimate, in the high-current condition, it is reasonable to assume that the electron beam waist on the particle is on the order of Ångstroms and the beam current approaches a nanoamp. These values suggest a current density of ~$10^6$ A/cm$^2$ and it is observed that the nanoparticle begins to undergo changes within minutes. This alteration is not entirely unexpected, as electron beam damage is a common problem in (S)TEM.[38–41] However, if properly monitored and carefully performed, the electron beam irradiation can lead to desirable structural and chemical modifications. In **Figure 2**, we show the time evolution of 'spot drilling' using a focused electron beam positioned in the center of an isolated nanoparticle existing outside of an array. Figures 2 (a) and (b) demonstrate that even after completely drilling through a single particle, crystallinity is preserved in the remainder of the nanoparticle. **Supplementary Figure S1** shows a drilling progression in time demonstrating that selective removal can approach single atomic column resolution. Meanwhile, Figure 2 (c) and (d) show the plasmonic responses evolving in time after irradiation. The energy maps in Figure 2(c) and spectral features in Figure 2(d) are extracted in an exploratory fashion using non-negative matrix factorization (NMF), a technique which has been applied for the analysis of EELS data.[42] A similar effect is observed after irradiation when selecting and viewing EEL spectra in plasmonic hot spots (e.g., corner, edge, bulk) and is shown in **Supplementary Figure S2**. The nanoparticle shape changes as observed in the HAADF-STEM images acquired every 30 seconds. Two observations are critical regarding the spatial structure of the plasmon response: first, the so-called "corner mode" in NMF component 1 develops into an all-around surface feature rather than being prominent only at the corners; second, a new feature manifests as a localized region in the particle center in NMF component 3 that initiates after irradiation for ~60 seconds. The bulk mode in NMF component 2 remains relatively unchanged, which indicates we have selectively altered the spatial distribution of specific plasmon modes, leaving some modes unaffected, and introduced a new region of electric field intensity within the nanoparticle. The three NMF components, in general, do not exhibit a change in their spectral peak position. Of critical importance here is the fact that the intensities of the modes do not decrease substantially after irradiation. While the NMF 1 spectral component decreases slightly, it can be argued this is due to the mode being spread out over the particle surface rather than concentrated in the corners. However, the bulk resonance (NMF 3) does decrease by about half, while the surface plasmon edge mode (NMF 2) essentially remains the same, suggesting the possibility of tailoring specific plasmon modes within a single structure.

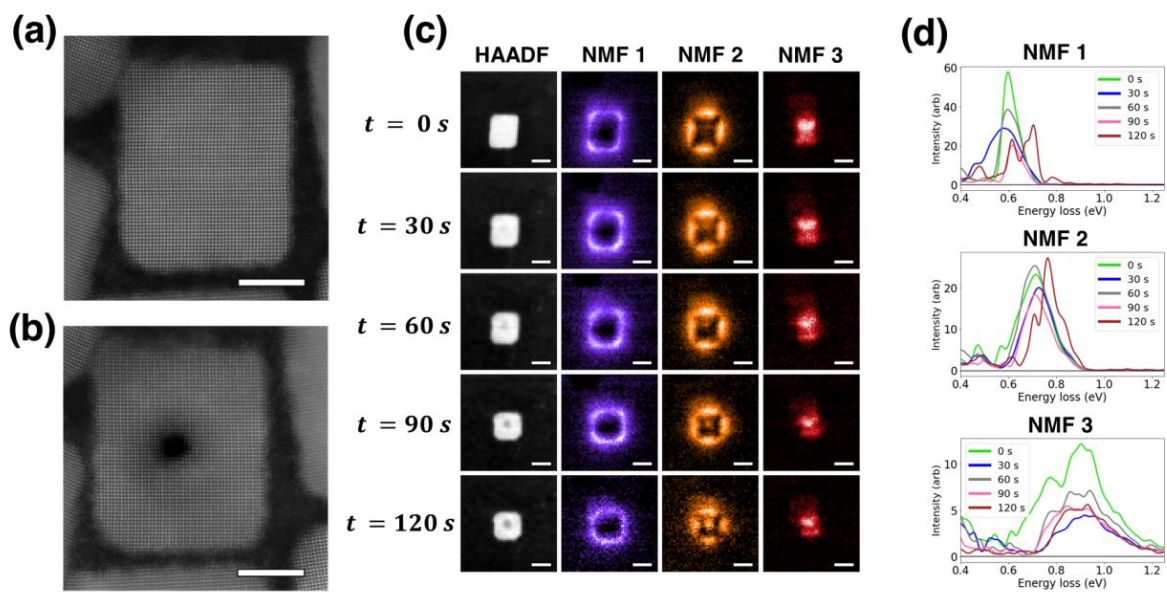

**Figure 2**. Spot drilling on an isolated nanoparticle. HAADF-STEM images acquired at low current showing crystal structure before drilling (a), and after drilling (b); large portion of nanoparticle retains crystallinity. HAADF-STEM images in (c) shown as a function of drilling center of nanoparticle, with NMF deconvolution performed in spectral domain to assess changes in plasmonic response due to spot drilling. Dominant three NMF abundance maps shown for each time step in (c), while corresponding NMF spectral endmembers, which are Gaussian blurred for clarity, shown in (d). Note that propensity of NMF method is used to separate different symmetries in spatial plasmonic maps; NMF endmembers illustrate evolution of corresponding behaviors during modification. Scale bar in (a) and (b) is 5 nm, all scale bars in (c) are 10 nm.

The fact that the crystallinity of the drilled particle (and consequently any nearby particle) remains intact is crucial for retaining intense plasmon features. A larger field of view showing the preservation of crystallinity in nearby particles is shown in **Supplementary Figure S3**. However, we also observe that once a hole has been drilled, the particle seems to be in a weakened state near the surface of hole, as even a lower current electron probe can enlarge the hole size with relative ease. We postulate that the atoms at the hole interface are left in a destabilized bonding state and therefore can be easily swept away by even a low intensity probe, though perhaps an *in-situ* or *ex-situ* annealing step can assist in forming a more strongly bonded inner hole surface after irradiation.

Nanoparticle removal can be performed using either a highly focused beam which scans a defined region, or a defocused beam fixed on the equivalent region – we note that the present authors have had more success with scanning a focused beam. It is also critical to note that a nanoparticle need not be completely drilled through or removed – instead, by reducing beam dwell time, one may choose to partially remove the material, in fact this can be seen in the HAADF-STEM images in both Figure 2 and Figure S1. However, this process still effectively modifies the shape, which in turn has an effect of the plasmon resonance.

Engineering the plasmon distribution in dimensions of both space and energy is also possible. We illustrate this effect by illuminating a particle dimer pair with a defocused probe whose beam diameter slightly exceeds that of a single nanoparticle. The probe is then placed between the two particles with the intention of fusing the particles into a single unit. This process is shown in discrete time intervals in **Figure 3**, where it is clear in the structural HAADF-STEM image that the particle shape is changing.

It is not surprising that the particle could become contaminated or rearrange its atomic configurations (e.g., reduced crystallinity), and therefore its plasmonic response would in turn become weaker. To study the exposure effect in a manner that allows spatial exploration of plasmonic activity, we perform NMF deconvolution on each dataset, and show the first and second NMF components at each time step. The NMF components show that the plasmon modes red-shift as a function of melding. Over the entire 3-min. exposure, we measure a 0.22 eV and a 0.14 eV redshift for components 1 and 2, respectively. In the **Supplementary Figure S4**, raw spectra from selected regions of interest are shown to strengthen the argument for a true redshift of the plasmon resonances. We again draw attention to the behavior that after electron irradiation, the surface resonances are more strongly affected spatially than the bulk resonance.

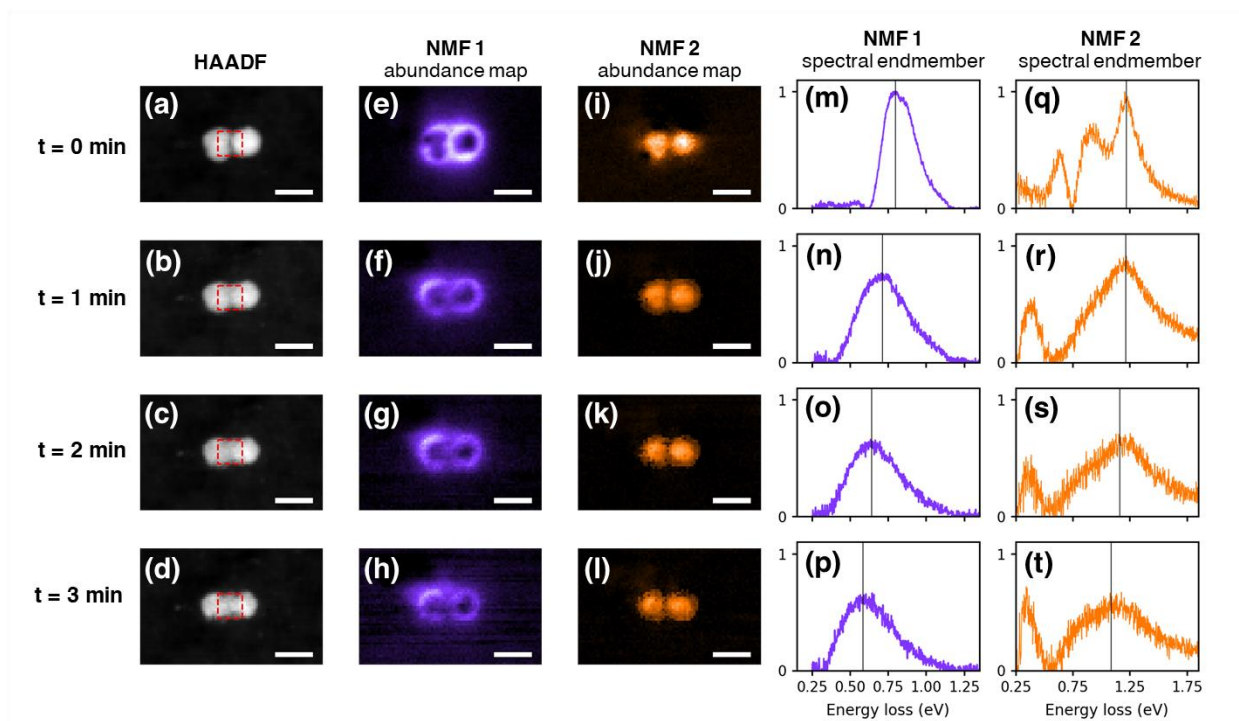

**Figure 3.** Effect of electron beam exposure on structure and plasmon resonance of a dimer configuration. HAADF-STEM images are shown in (a-d); 1st NMF and 2nd NMF abundance maps for each time increment are shown in (e-h) and (i-l), respectively; 1st NMF and 2nd NMF spectral endmembers for each time increment are shown in (m-p) and (q-t), respectively, normalized to strongest mode. Beam was defocused and placed in region between nanoparticles to promote fusing, shown by the dashed region in the HAADF-STEM images. Black vertical lines in (m-t) indicate spectral peak position of dominant feature for plasmon mode of interest. All scale bars are 20 nm.

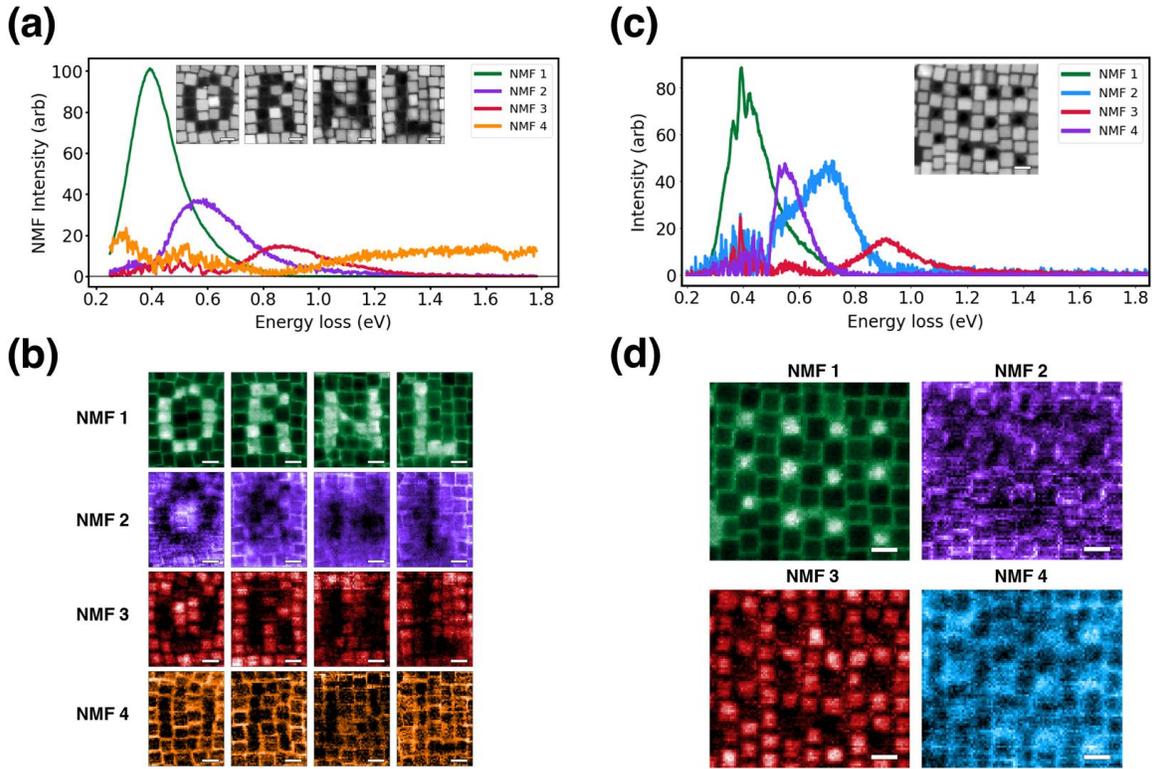

**Figure 4**. Examples of spatial modification of plasmon responses. (a) and (b) show laboratory logo "ORNL" shaped by electron beam; four dominant NMF spectral features with corresponding spatial maps are shown in (a) and (b) respectively; (c) and (d) depict periodic control in the form of a checkerboard pattern and dominant NMF components emphasize resulting spatial localization of different plasmon modes. HAADF-STEM images of each pattern are insets in (a) and (c). Scale bars are all 20 nm.

We also utilize the self-assembled array as a basis to create more complex structures since the array supports a wide range of plasmon modes that represent collective responses of the entire system. In the vicinity of a defect such as an edge or missing particle within the array, the plasmonic response extends into this space with high intensity. This effect is crucial for many applications that require an enhanced field strength in a specific location that, as we will show, can be achieved by manipulation of a nanoparticle's position using the electron beam. To illustrate the spatial modification to the plasmonic features, in **Figure 4** we selectively remove particles to produce specifically designed structures, e.g., the ORNL logo in Figure 4 (a), or a checkerboard type pattern in Figure 4 (b). The plasmonic effects of other geometries shown in Figure 1 are presented in **Supplementary Figure S5**. The result is strong plasmonic intensity at selectable locations with nanoscale resolution, which is induced by selectively removing nanoparticles from the array. The NMF components highlight the different regions of plasmonic activity; and their associated spectral features are also shown in color.

Perhaps not immediately obvious but nonetheless notable is the fact that the resonances are controlled in several different ways, for example, by removing a particle a void is created where the plasmonic intensity is still strongly observed. At the same time, since the particle is missing, there is nothing to support a bulk mode at that location. This change in response is illustrated by the various components' spatial maps in Figure 4. In Figure 4 (a), NMF 1 (green) shows the collective plasmon mode that normally is evenly distributed across a regular array. However, due to the creation of defects (i.e., removal of particles in the array), we find that the delocalized surface plasmon array mode extends into these defects, allowing us to create localized plasmonic voids where the array mode is still strongly activated, but the localized surface and bulk modes are no longer present. In other words, this enables us to modify the overall structure of the plasmon response. A careful examination of the distance dependence of the plasmonic intensity is illustrated in **Supplementary Figure S6**. Presenting the response as shown in Figure 4 merely serves to emphasize the spatial control of specific energy modes. Similar effects are shown for the checkerboard pattern in Figure 4 (b), where emphasis is placed on the ability to generate periodic spatial structures. The resulting localized plasmon resonances in these periodic checkerboards are spectrally similar, but their locations in space are controlled.

Irradiation of a material by the electron beam can involve several processes[43] dependent on the electron beam energy, the probe current, the material system, the support, etc. In an effort to determine the primary mechanism responsible for the deliberate changes, we employed several different accelerating voltages. The substrate, a 15-nm thick silicon nitride membrane, remains the same in all cases. We did not observe a strong dependence on beam energy (60-200 kV) with the beam current kept constant at ~0.5 nA for particle removal; however, these experiments show a clear dependence on current, i.e., increasing probe current hastens particle removal. A detailed understanding of the mechanism behind the nanoparticle removal by the electron beam is not the intent of this work, however, we are able to trace the chemical evolution, which offers insight into the origins of the plasmonic feature modifications.

We explore the chemical and physical changes occurring in individual nanoparticles as well as nearby nanoparticles. For nanoparticles that are completely or partially removed, a primary concern is redeposition of the removed material and general contamination of nearby nanoparticles. From an EELS analysis of both low loss and core loss energy signals, we show in **Supplementary Figure S7** that the chemical composition of nearby particles is preserved. We speculate that the material may be entirely removed from the system since we also observe the removal of the underlying silicon nitride membrane. The substrate removal was a little surprising, as Ref [44] suggests a minimum beam energy of 150 kV is required to drill through amorphous silicon nitride, but this is likely due to the effect of our high current density for drilling and ultra-high-vacuum conditions. Otherwise, if partially drilling a particle, the removed material will be in an amorphous state and can migrate around the surface of the local sample region based on the probe location. We also note the presence of a small amount of carbon contamination that appears to coat all nanoparticles uniformly, but this is present even in regular imaging conditions far from any regions that were sculpted.

To address the question of which elemental species are removed and where they might go, we perform time-resolved EELS, which are shown as part of the supplementary information. For nanoparticles that are partially drilled, or for those that are fused together, this time dependent analysis is performed by capturing the local core loss EELS signal during the irradiation process. In **Supplementary Figure S8**, the time dependence of the core loss EELS in 20 s intervals is shown, where the EEL spectrum has been acquired every 1 s and integrated over 20 s. From Figure S8 (a), the indium oxide features at 532 and 540 (~550-560) eV weaken significantly over the course of 3 min., while the oxygen K edge remains essentially unchanged. The broad indium $M_{4,5}$ edge near 443 eV as well as the weaker $M_3$ indium edge at 664 eV vanish nearly completely during the total exposure, which indicates preferential removal of the metallic species from the localized region. Notably, both indium and tin have very low melting points. It is certainly possible we are removing the F alternatively, as this is a dopant and tends to be easily removed by electron beam, and also may explain the red-shifting behavior due to lack of electron donation. Unfortunately, the core loss EEL signature of Fluorine is difficult to detect. In any case, the preferential removal of different elements might present opportunities for the direct chemical engineering of atomic structures; as well as an enhanced scope for plasmon engineering via local control of the conduction electron levels.

Considering the observation that the indium and fluorine content is modified suggests that the conduction electrons supporting the plasmon resonances will be decreased due to reduced metallic character in the nanocrystals. Since we have already shown that a majority of the crystallinity is preserved, redeposition of removed material into an amorphous state covering the particles can explain the partially *reduced* plasmon intensity and is in general not overly surprising. We also expect the beam to attract carbon contamination onto the surface of the nanocubes, further dampening the plasmon response. What is peculiar and of great potential utility is the spectral shifting of the plasmon signals after irradiation. Spectral tunability in metal nanoparticles is generally accomplished via particle size and shape tuning – i.e., larger and elongated objects support longer wavelengths, however it is also possible to tune plasmon resonances by adjusting the number of conduction electrons in the material, for example by doping[34]. Here, we instead altered the concentration of conduction electrons in the material by using the electron beam, and also have the ability to modify the shape; thus, this work presents two nanoscale methods for tuning the resonant frequencies.

Future studies will examine these effects for electron beam irradiation in other metal oxide plasmonic and noble metal nanoparticles. We postulate that it is the more complex chemical structure of the oxides here that allows such a chemical modification of the plasmonic structure, and attempts to repeat similar experiments in noble metal nanoparticles might produce different results. We would likely not expect to modify the spectral position of the resonances in single-element nanoparticles in the same way. As such, the average concentration of conduction electrons would not change upon removal of small amounts of noble metal atoms, and for this reason, we suspect this chemical effect has not been previously reported.

In summary, we have shown that sculpting of the plasmonic properties of nanoparticles with a resolution better than 1 nm is possible using a high-energy focused electron beam under appropriate conditions. A great degree of flexibility is enabled, allowing particles to be either partially removed, completely removed, fused together, and, if desired, selectively modified in terms of chemistry and crystal structure. Precise removal of nanoparticles enables the design of increasingly complex nanophotonic and plasmonic systems by dynamically modifying as-assembled structures on-the-fly. Critically, the plasmonic responses of the nanoparticles were modified in the spatial domain, producing unique plasmonic systems, which carry a corresponding change in their energy response. Irradiation by an electron beam was demonstrated to have different effects depending on critical factors, such as electron dose imparted to a nanoparticle, area of irradiation, amount of defocus, etc. Importantly, chemical and plasmonic analysis can be carried out seamlessly by EELS at different energy scales immediately following the electron beam modification.

Moving on from tailoring the spatial response, we envision an automated experiment where a particular plasmonic feature can be specifically designed and subsequently engineered by precise automated control and feedback of the electron beam. Recent work using encoder-decoder neural networks[45] demonstrated extraction of the correlative relationship between the local particle geometry and plasmonic response. Hence, we aim to generate custom responses where the network predicts the required geometry and then, as shown here, we deliberately sculpt such a geometry using e-beam particle modification methods to match that template. The future automation of this experiment will allow a deeper exploration into fully customizable and almost arbitrary plasmonic behavior, and process automation will lead to a significantly faster, accurate, and, perhaps most important of all, reproducible results.

**Acknowledgements:** This effort is based upon work supported by the U.S. Department of Energy (DOE), Office of Science, Basic Energy Sciences (BES), Materials Sciences and Engineering Division (K.M.R., S.V.K.) S.H.C and D.J.M. acknowledge (NSF CHE-1905263, and CDCM, an NSF MRSEC DMR-1720595), the Welch Foundation (F-1848), and the Fulbright Program (IIE-15151071). Electron microscopy was performed using instrumentation within ORNL's Materials Characterization Core provided by UT-Battelle, LLC, under Contract No. DE-AC05- 00OR22725 with the DOE and sponsored by the Laboratory Directed Research and Development Program of Oak Ridge National Laboratory, managed by UT-Battelle, LLC, for the U.S. Department of Energy.

**Author contributions:** K.M.R. conducted the microscopy, performed analysis, and wrote the manuscript. S.H.C synthesized the samples and help write the manuscript. A.R.L enabled the optimal beam control, helped consider additional experimental methods and helped write the manuscript. D.J.M. helped write the manuscript. S.V.K. conceived and oversaw the project, and helped write the manuscript. We thank Jordan Hachtel for his insightful discussions and invaluable advice.

**The authors declare no competing interests.**

# References

1. Alivisatos, A. P. Semiconductor Clusters, Nanocrystals, and Quantum Dots. *Science* **271**, 933–937 (1996).

2. Kovalenko, M. V. *et al.* Prospects of Nanoscience with Nanocrystals. *ACS Nano* **9**, 1012–1057 (2015).

3. Talapin, D. V., Lee, J.-S., Kovalenko, M. V. & Shevchenko, E. V. Prospects of Colloidal Nanocrystals for Electronic and Optoelectronic Applications. *Chem. Rev.* **110**, 389–458 (2010).

4. Michalet, X. *et al.* Quantum Dots for Live Cells, in Vivo Imaging, and Diagnostics. *Science* **307**, 538–544 (2005).

5. Fan, J. A. *et al.* Self-Assembled Plasmonic Nanoparticle Clusters. *Science* **328**, 1135–1138 (2010).

6. Yang, P., Zheng, J., Xu, Y., Zhang, Q. & Jiang, L. Colloidal Synthesis and Applications of Plasmonic Metal Nanoparticles. *Advanced Materials* **28**, 10508–10517 (2016).

7. Agrawal, A. *et al.* Localized Surface Plasmon Resonance in Semiconductor Nanocrystals. *Chem. Rev.* **118**, 3121–3207 (2018).

8. Tame, M. S. *et al.* Quantum plasmonics. *Nature Physics* **9**, 329–340 (2013).

9. Corma, A. & Garcia, H. Supported gold nanoparticles as catalysts for organic reactions. *Chemical Society Reviews* **37**, 2096–2126 (2008).

10. Lee, D., Rubner, M. F. & Cohen, R. E. All-Nanoparticle Thin-Film Coatings. *Nano Lett.* **6**, 2305–2312 (2006).

11. Bobo, D., Robinson, K. J., Islam, J., Thurecht, K. J. & Corrie, S. R. Nanoparticle-Based Medicines: A Review of FDA-Approved Materials and Clinical Trials to Date. *Pharm Res* **33**, 2373–2387 (2016).

12. Jain, P. K., Huang, X., El-Sayed, I. H. & El-Sayed, M. A. Review of Some Interesting Surface Plasmon Resonance-enhanced Properties of Noble Metal Nanoparticles and Their Applications to Biosystems. *Plasmonics* **2**, 107–118 (2007).

13. Otto, A., Grabhorn, H. & Akemann, W. Surface-enhanced Raman scattering. *J. Phys.: Condes. Matter* **4**, 1143–1212 (1992).

14. Aćimović, S. S., Kreuzer, M. P., González, M. U. & Quidant, R. Plasmon Near-Field Coupling in Metal Dimers as a Step toward Single-Molecule Sensing. *ACS Nano* **3**, 1231–1237 (2009).

15. Wang, H., Brandl, D. W., Nordlander, P. & Halas, N. J. Plasmonic Nanostructures: Artificial Molecules. *Acc. Chem. Res.* **40**, 53–62 (2007).

16. Lee, J.-H. *et al.* Artificially engineered magnetic nanoparticles for ultra-sensitive molecular imaging. *Nature Medicine* **13**, 95–99 (2007).

17. Milliron, D. J. *et al.* Colloidal nanocrystal heterostructures with linear and branched topology. *Nature* **430**, 190–195 (2004).

18. Grzelczak, M., Pérez-Juste, J., Mulvaney, P. & Liz-Marzán, L. M. Shape control in gold nanoparticle synthesis. *Chem. Soc. Rev.* **37**, 1783–1791 (2008).

19. Pandey, P. A. *et al.* Physical Vapor Deposition of Metal Nanoparticles on Chemically Modified Graphene: Observations on Metal–Graphene Interactions. *Small* **7**, 3202–3210 (2011).

20. Shevchenko, E. V., Talapin, D. V., Kotov, N. A., O'Brien, S. & Murray, C. B. Structural diversity in binary nanoparticle superlattices. *Nature* **439**, 55–59 (2006).

21. Fowlkes, J. D., Kondic, L., Diez, J., Wu, Y. & Rack, P. D. Self-Assembly versus Directed Assembly of Nanoparticles via Pulsed Laser Induced Dewetting of Patterned Metal Films. *Nano Lett.* **11**, 2478–2485 (2011).

22. Randolph, S. J., Fowlkes, J. D. & Rack, P. D. Focused, Nanoscale Electron-Beam-Induced Deposition and Etching. *Critical Reviews in Solid State and Materials Sciences* **31**, 55–89 (2006).

23. Boles, M. A., Engel, M. & Talapin, D. V. Self-Assembly of Colloidal Nanocrystals: From Intricate Structures to Functional Materials. *Chem. Rev.* **116**, 11220–11289 (2016).

24. Manfrinato, V. R. *et al.* Aberration-Corrected Electron Beam Lithography at the One Nanometer Length Scale. *Nano Lett.* **17**, 4562–4567 (2017).

25. Egerton, R. F. Control of radiation damage in the TEM. *Ultramicroscopy* **127**, 100–108 (2013).

26. Dyck, O., Kim, S., Kalinin, S. V. & Jesse, S. Placing single atoms in graphene with a scanning transmission electron microscope. *Appl. Phys. Lett.* **111**, 113104 (2017).

27. Dyck, O. *et al.* Building Structures Atom by Atom via Electron Beam Manipulation. *Small* **14**, 1801771 (2018).

28. Susi, T. *et al.* Silicon--Carbon Bond Inversions Driven by 60-keV Electrons in Graphene. *Phys. Rev. Lett.* **113**, 115501 (2014).

29. Hudak, B. M. *et al.* Directed Atom-by-Atom Assembly of Dopants in Silicon. *ACS Nano* **12**, 5873–5879 (2018).

30. Tizei, L. H. G. *et al.* Tailored Nanoscale Plasmon-Enhanced Vibrational Electron Spectroscopy. *Nano Lett.* **20**, 2973–2979 (2020).

31. Cho, S. H. *et al.* Syntheses of Colloidal F:In2O3 Cubes: Fluorine-Induced Faceting and Infrared Plasmonic Response. *Chem. Mater.* **31**, 2661–2676 (2019).

32. Scholl, J. A., Koh, A. L. & Dionne, J. A. Quantum plasmon resonances of individual metallic nanoparticles. *Nature* **483**, 421–427 (2012).

33. Hachtel, J. A. *et al.* Identification of site-specific isotopic labels by vibrational spectroscopy in the electron microscope. *Science* **363**, 525–528 (2019).

34. Cho, S. H. *et al.* Spectrally tunable infrared plasmonic F,Sn:In2O3 nanocrystal cubes. *J. Chem. Phys.* **152**, 014709 (2020).

35. Staller, C. M., Gibbs, S. L., Saez Cabezas, C. A. & Milliron, D. J. Quantitative Analysis of Extinction Coefficients of Tin-Doped Indium Oxide Nanocrystal Ensembles. *Nano Lett.* **19**, 8149–8154 (2019).

36. Gibbs, S. L. *et al.* Intrinsic Optical and Electronic Properties from Quantitative Analysis of Plasmonic Semiconductor Nanocrystal Ensemble Optical Extinction. *J. Phys. Chem. C* **124**, 24351–24360 (2020).

37. Kalinin, S. V. *et al.* Separating physically distinct mechanisms in complex infrared plasmonic nanostructures via machine learning enhanced electron energy loss spectroscopy. *arXiv:2009.08501 [cond-mat]* (2020).

38. Koh, A. L. *et al.* Electron Energy-Loss Spectroscopy (EELS) of Surface Plasmons in Single Silver Nanoparticles and Dimers: Influence of Beam Damage and Mapping of Dark Modes. *ACS Nano* **3**, 3015–3022 (2009).

39. Hwang, S., Han, C. W., Venkatakrishnan, S. V., Bouman, C. A. & Ortalan, V. Towards the low-dose characterization of beam sensitive nanostructures via implementation of sparse image acquisition in scanning transmission electron microscopy. *Meas. Sci. Technol.* **28**, 045402 (2017).

40. Gonzalez-Martinez, I. G. *et al.* Electron-beam induced synthesis of nanostructures: a review. *Nanoscale* **8**, 11340–11362 (2016).

41. Jiang, N. Electron beam damage in oxides: a review. *Rep. Prog. Phys.* **79**, 016501 (2015).

42. Shiga, M. *et al.* Sparse modeling of EELS and EDX spectral imaging data by nonnegative matrix factorization. *Ultramicroscopy* **170**, 43–59 (2016).

43. Egerton, R. F., Li, P. & Malac, M. Radiation damage in the TEM and SEM. *Micron* **35**, 399–409 (2004).

44. Howitt, D. G., Chen, S. J., Gierhart, B. C., Smith, R. L. & Collins, S. D. The electron beam hole drilling of silicon nitride thin films. *Journal of Applied Physics* **103**, 024310 (2008).

45. Roccapriore, K. M., Ziatdinov, M., Cho, S. H., Hachtel, J. A. & Kalinin, S. V. Predictability of localized plasmonic responses in nanoparticle assemblies. *Small* (2021) doi: 10.1002/smll.202100181.

**Supplementary Information**

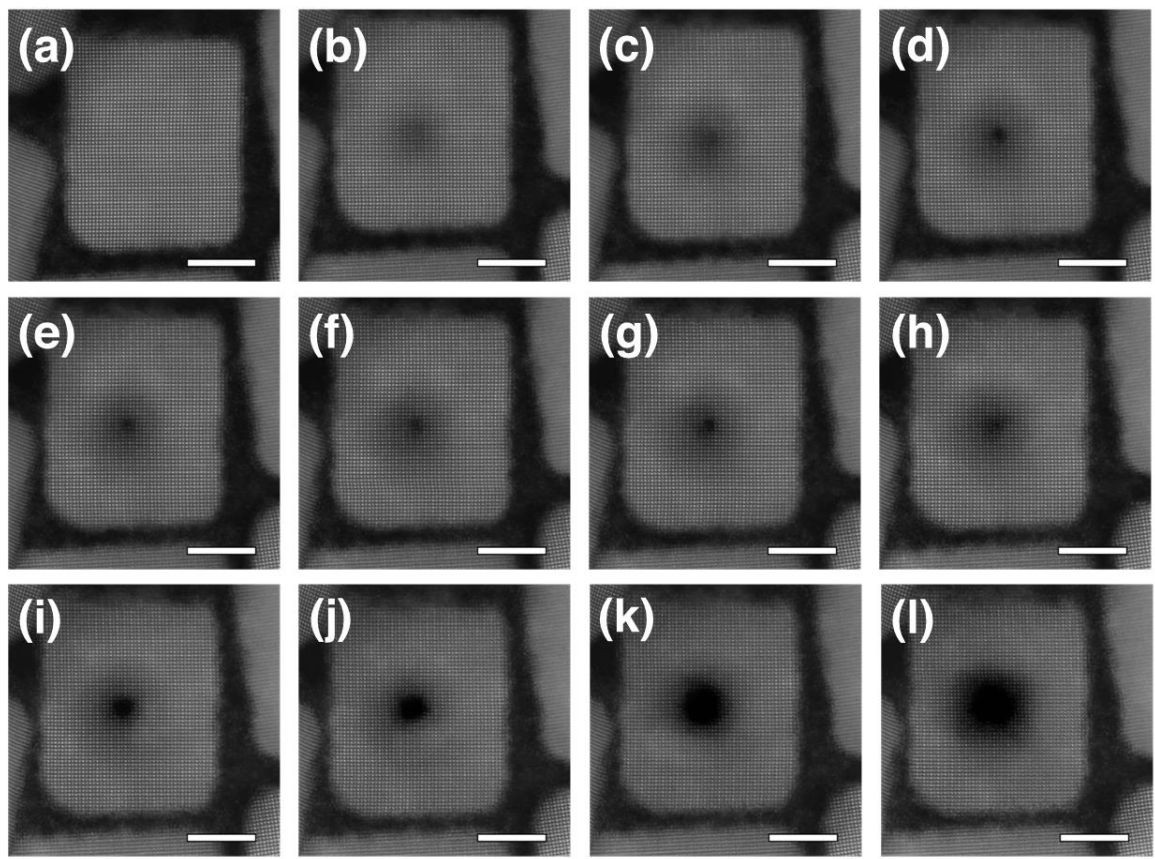

**Figure S1**. HAADF-STEM image time series of spot-drilling single nanoparticle. Progression of drilling in time steps of 30 seconds at reduced current compared to faster higher current drilling shown in (a-l), where atomic contrast imaging is immediately allowed following a drill step. All scale bars are 5 nm.

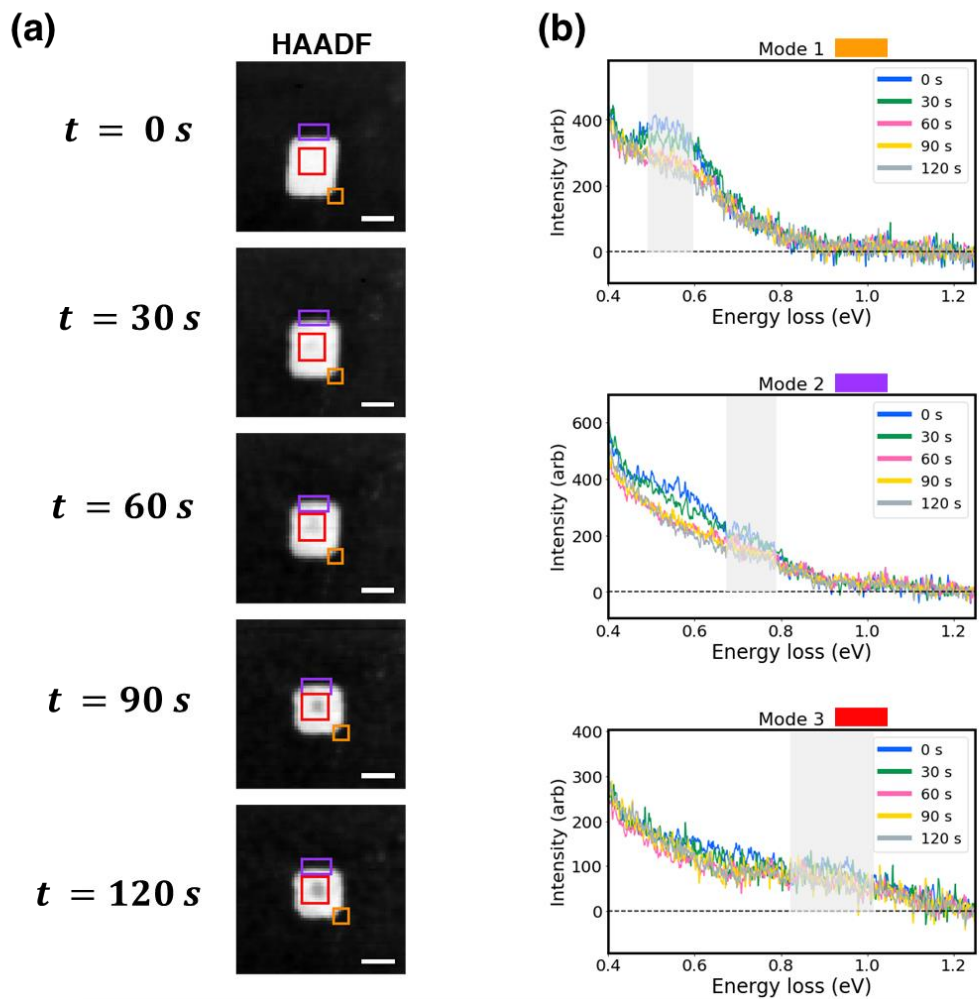

**Figure S2**. EEL spectra at selected ROIs near nanoparticle. (a) shows HAADF-STEM images of particle in time steps of 30 seconds while (b) shows spectra at locations shown in colored rectangles in (a) as a function of time; orange (corner) is mode 1, purple (edge) is mode 2, red (bulk) is mode 3. Transparent rectangles on (b) denote spectral region of interest for particular plasmonic feature, as multple modes can be spatially degenerate. All spectra are background subtracted to remove ZLP tails. Scale bars in (a) are 10 nm.

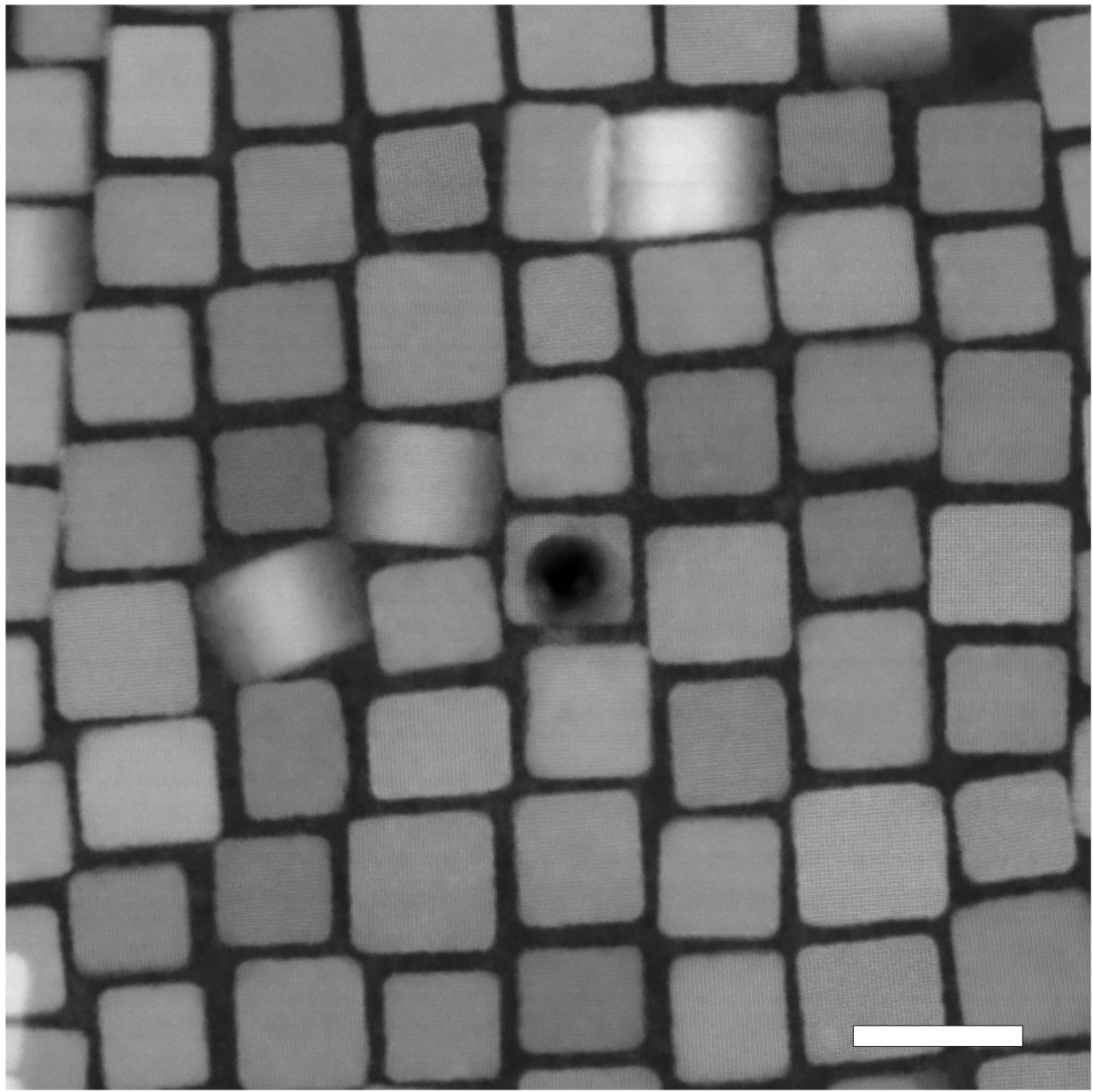

**Figure S3**. Preservation of crystallinity of drilled nanoparticle and nearby particles. Redeposited material is not observed on nearby particles, but is observed near bottom of drilled particle. Note that not all particles are on zone axis, therefore atomic contrast is not visible for all. Scale bar is 20 nm.

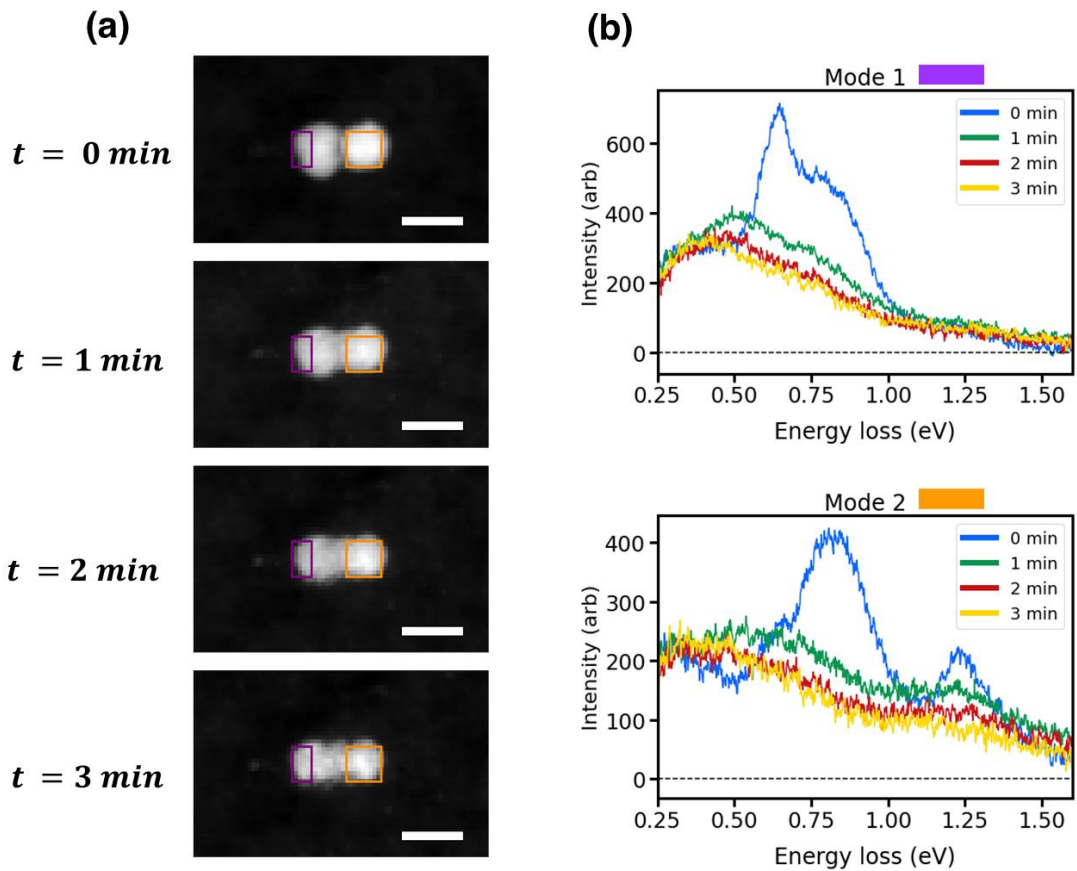

**Figure S4**. EEL spectra at selected ROIs near fused nanoparticle pair. (a) shows HAADF-STEM images of dimer in time steps of 1 min. while (b) shows spectra at locations shown in colored rectangles in (a) as a function of time; purple (edge) is mode 1, orange (bulk) is mode 2. Note spectra are background subtracted to remove effect of ZLP tails. Scale bars in (a) are 20 nm.

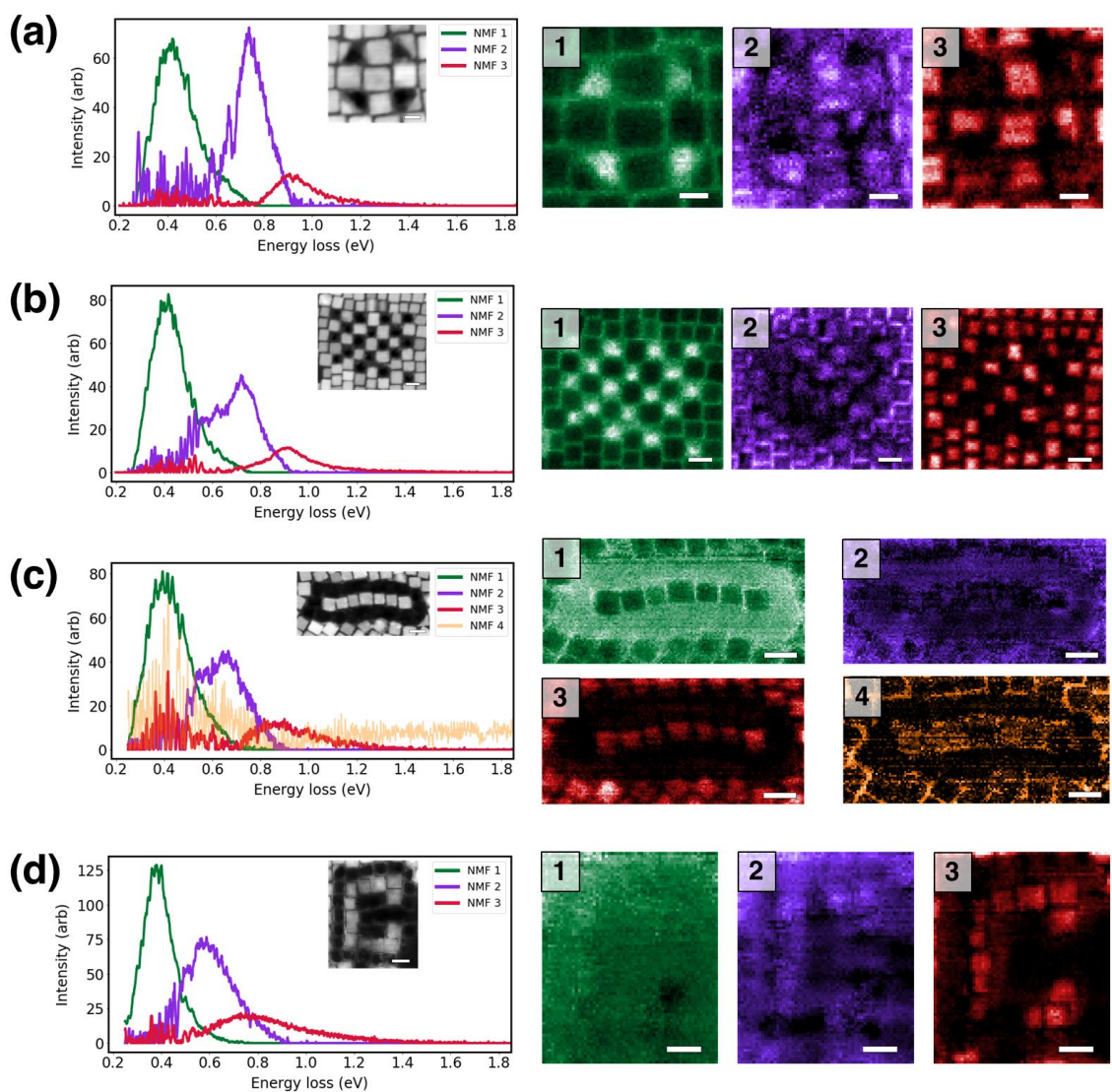

**Figure S5**. Additonal sculpted configurations, with NMF spectral features (left) and corresponding NMF energy maps (right). (a) diagonally cut half-nanoparticles, (b) another checkerboard with more complex periodicity, (c) One-dimensional chain of nanocubes, and (d) split-ring resonator. Note that NMF mode 3 (bulk mode) does not appear in diagonal half particles in (c), and NMF 1 in (d) is no longer localized. Due to significant material removal, there was considerable redeposition of ejected nanoparticle material and carbon contamination on (c) and (d), hence their weak spectral output. Scale bar in (a) is 10 nm, and 20 nm in (b-d).

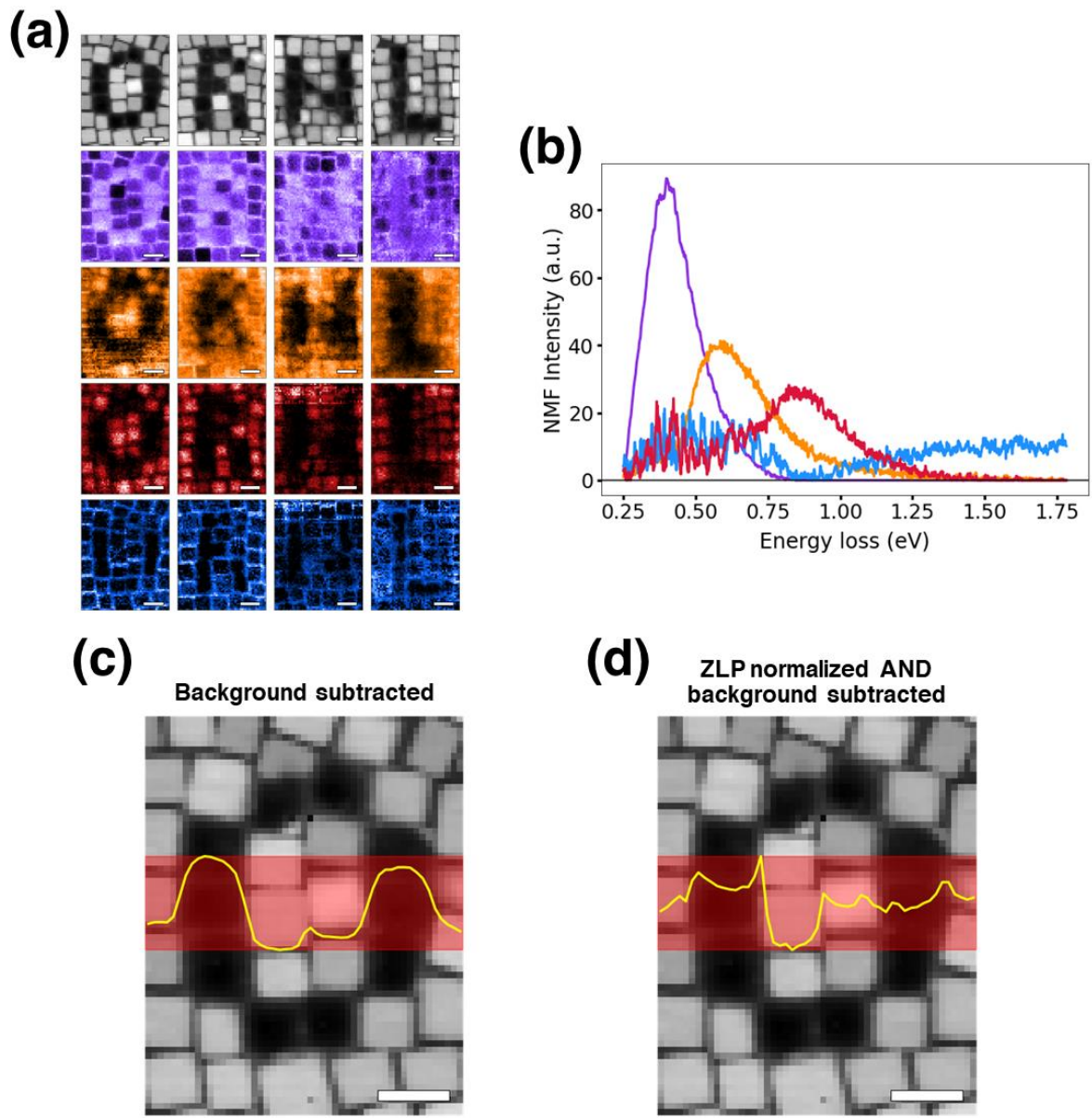

**Figure S6**. Effect of normaling to ZLP to account for elastic scattering. (a) NMF component maps of ORNL logo and corresponding spectral endmembers (b) after ZLP normalization. Line profiles taken across shaded regions in (c) and (d), displaying maximum pixel intensity as function of pixel position in horizontal direction. Note that possible enhancement and spatial control of plasmon intensity is qualitatively the same as shown in Figure 3. All scale bars 20 nm.

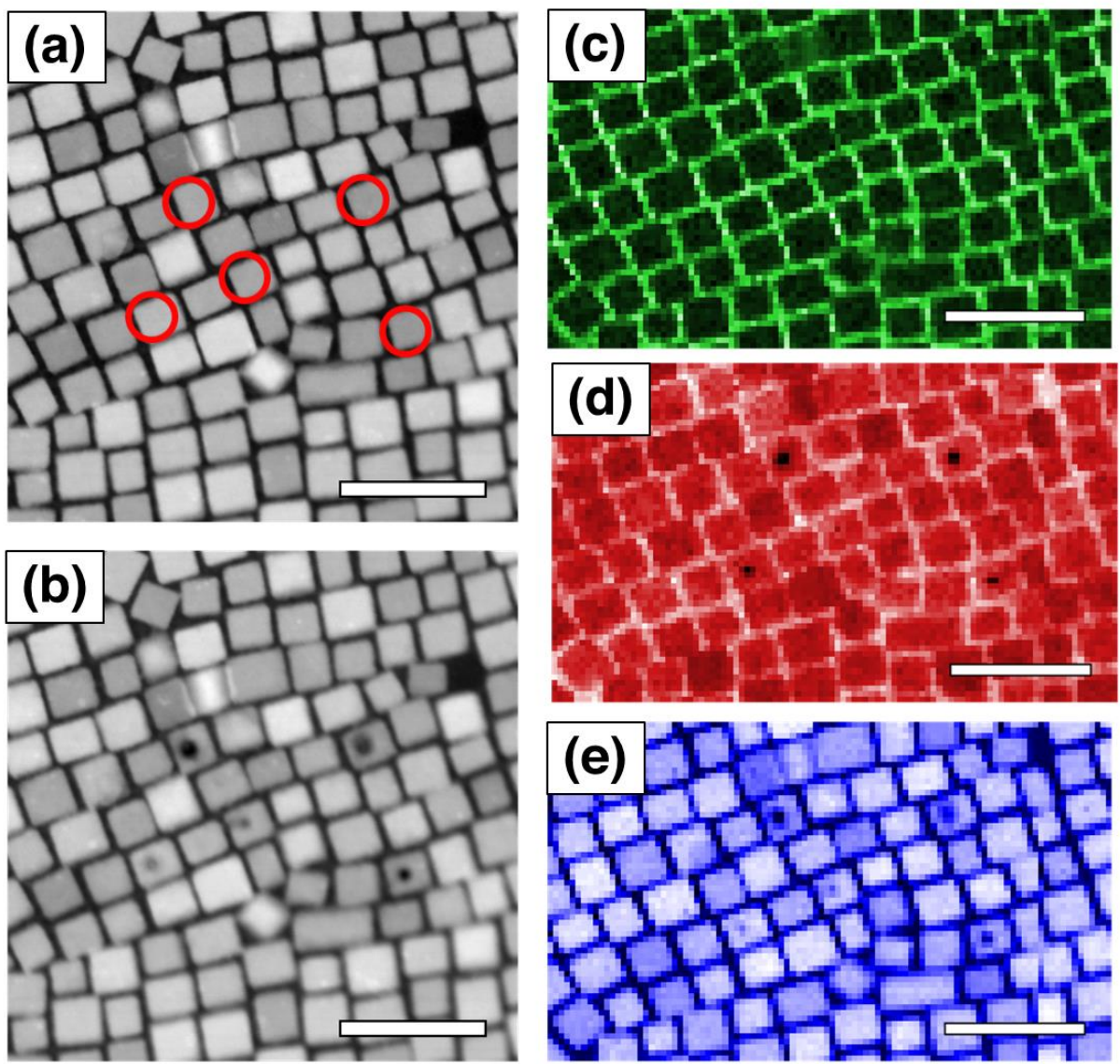

**Figure S7**. Chemical changes from electron irradiation. The set of HAADF-STEM images in (a) and (b) show region before and after exposure, respectively, of five different particles at varying dwell times, identified by red circles in (a). Integrated energy regions for Carbon K edge (284-334 eV), Nitrogen K edge (401-431 eV), and indium oxide (532-562 eV) features are shown in (c), (d), and (e) respectively. Chemistry of nanoparticles not intentionally exposed remains preserved. Scale bars are all 50 nm.

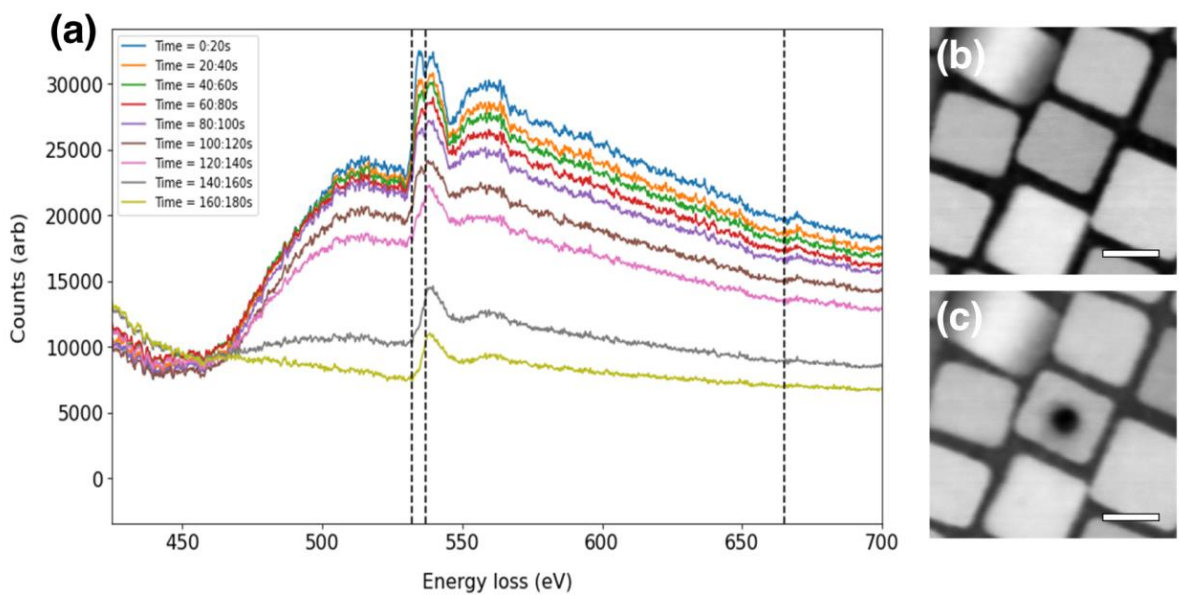

**Figure S8**. Time evolution of core loss EEL spectra for chemical analysis. In (a), the core-loss EELS In$_2$O$_3$ signature (~550 eV) in general fades in time as electron beam is fixed on particle; however, oxygen signal remains relatively unaffected, while the minor M$_3$ indium edge at 664 eV vanishes, indicating preferential reduction of metallic character (b) and (c) show the HAADF-STEM images before and after irradiation. Scale bars in (b) and (c) 10 nm.